\documentclass[12pt, a4paper,fleqn]{article}
\pdfoutput=1
\usepackage{natbib}
\usepackage[only,fivrm]{rawfonts}
\usepackage{uebungsk}
\usepackage{url}
\usepackage{bbm}
\usepackage{tabularx}
\usepackage{dcolumn}
\usepackage{color}
\usepackage{booktabs}
\usepackage{graphicx,rotating,epsfig,multirow,multicol,hhline}
\usepackage{amsmath,amsthm,amssymb,amsfonts}
\usepackage{rotating}
\usepackage{meinfuss}
\usepackage{slashbox}
\usepackage{float}
\usepackage[hang]{caption2}
\usepackage{textcomp}
\usepackage[latin1]{inputenc}
\renewcommand{\baselinestretch}{1.5}

\newcommand{\adv}{{\rm AE}}

\newcommand{\p}{{\rm P}}

\bibpunct{[}{]}{,}{n}{}{,}

\begin{document}

\begin{titlepage}
\vspace{-1.0cm}
\begin{center}
\renewcommand{\baselinestretch}{1.0}
\Large{\textbf{On estimands and the analysis of adverse events in the presence of varying follow-up times within the benefit assessment of therapies}} \\
\vspace{0.5cm}
\renewcommand{\thefootnote}{\fnsymbol{footnote}}
\normalsize{S. Unkel$^{1}$\footnote[1]{Correspondence should be
addressed to: Steffen Unkel, Department of Medical Statistics, University Medical Center Goettingen, Humboldtallee 32, 37073 Goettingen, Germany (\url{e-mail: steffen.unkel@med.uni-goettingen.de}).}, M. Amiri$^{2}$, N. Benda$^{3}$, J. Beyersmann$^{4}$, D. Knoerzer$^{5}$, \newline K. Kupas$^{6}$, F. Langer$^{7}$, F. Leverkus$^{8}$, A. Loos$^{9}$, C. Ose$^{2}$, T. Proctor$^{10}$,\newline C. Schmoor$^{11}$, C. Schwenke$^{12}$, G. Skipka$^{13}$, K. Unnebrink$^{14}$, F. Voss$^{15}$, and T. Friede$^{1}$}  \\
\vspace{0.5cm}
\renewcommand{\baselinestretch}{1.0}
$^{1}$ \small{Department of Medical Statistics, University Medical Center Goettingen, Germany} \\
$ $ \newline
$^{2}$ \small{Center for Clinical Trials, University Hospital Essen, Germany} \\
$ $ \newline
$^{3}$ \small{Biostatistics and Special Pharmacokinetics Unit,} \\
\small{Federal Institute for Drugs and Medical Devices, Bonn, Germany} \\
$ $ \newline
$^{4}$ \small{Institute of Statistics, Ulm University, Germany} \\
$ $ \newline
$^{5}$ \small{Roche Pharma AG, Grenzach, Germany} \\
$ $ \newline
$^{6}$ \small{Bristol-Myers Squibb GmbH \& Co. KGaA, München, Germany} \\
$ $ \newline
$^{7}$ \small{Lilly Deutschland GmbH, Bad Homburg, Germany} \\
$ $ \newline
$^{8}$ \small{Pfizer Deutschland GmbH, Berlin, Germany} \\
$ $ \newline
$^{9}$ \small{Merck KGaA, Darmstadt, Germany} \\
$ $ \newline
$^{10}$ \small{Institute of Medical Biometry and Informatics,} \\
\small{University of Heidelberg, Germany} \\
$ $ \newline
$^{11}$ \small{Clinical Trials Unit, Faculty of Medicine and Medical Center - University of Freiburg, Germany} \\
$ $ \newline
$^{12}$ \small{Schwenke Consulting: Strategies and Solutions in Statistics (SCO:SSIS),} \\
\small{Berlin, Germany} \\
$ $ \newline
$^{13}$ \small{Institute for Quality and Efficiency in Health Care, Cologne, Germany} \\
$ $ \newline
$^{14}$ \small{AbbVie Deutschland GmbH \& Co. KG, Ludwigshafen, Germany} \\
$ $ \newline
$^{15}$ \small{Boehringer Ingelheim Pharma GmbH \& Co. KG, Ingelheim, Germany} \\
\today
\end{center}
\pagebreak
\thispagestyle{empty}
\begin{abstract}
\renewcommand{\baselinestretch}{1.5}
\renewcommand{\thefootnote}{\fnsymbol{footnote}}
The analysis of adverse events (AEs) is a key component in the assessment of a drug's safety profile. Inappropriate analysis methods may result in misleading conclusions about a therapy's safety and consequently its benefit-risk ratio. The statistical analysis of AEs is complicated by the fact that the follow-up times can vary between the patients included in a clinical trial. This paper takes as its focus the analysis of AE data in the presence of varying follow-up times within the benefit assessment of therapeutic interventions. Instead of approaching this issue directly and solely from an analysis point of view, we first discuss what should be estimated in the context of safety data, leading to the concept of estimands. Although the current discussion on estimands is mainly related to efficacy evaluation, the concept is applicable to safety endpoints as well. Within the framework of estimands, we present statistical methods for analysing AEs with the focus being on the time to the occurrence of the first AE of a specific type. We give recommendations which estimators should be used for the estimands described. Furthermore, we state practical implications of the analysis of AEs in clinical trials and give an overview of examples across different indications. We also provide a review of current practices of health technology assessment (HTA) agencies with respect to the evaluation of safety data. Finally, we describe problems with meta-analyses of AE data and sketch possible solutions.
\end{abstract}

\renewcommand{\baselinestretch}{1.5}


\end{titlepage}
\section{Setting the scene}
 The analysis of adverse events (AEs) is a key component in the assessment of a drug's safety profile. Inappropriate analysis methods may result in misleading conclusions about a therapy's safety and consequently its benefit-risk ratio. A variety of methods are available for the analysis of AEs, but their complexity and the imposed assumptions clearly differ. The simplest methods for contingency tables \citep{Agresti2013}, e.g. na\"{\i}ve proportions and derived effect measures such as risk differences, relative risks and odds ratios, presuppose identical follow-up times and usually ignore recurrent events and competing risks. 
If follow-up times are different across treatment groups, then comparisons based on simple incidence proportions produce biased results. Consideration of varying follow-up times by means of incidence densities is possible. However, the incidence density relies on a rather restrictive constant hazard assumption. Standard procedures for event times \citep[see e.g.][]{Collett2015a}, in turn, are based on non-informative censoring and are not readily suitable for recurrent events and competing risks. For this purpose, more complex methods in the area of event time analysis do exist \citep[see e.g.][Chapters 12--13]{Collett2015a}; for whose application, however, the data must in turn meet the corresponding prerequisites.\newline
The purpose of the present paper is to address the research gap in the analysis of AE data in the spirit of the current discussion on clinical trial estimands. Instead of approaching the problem of investigating AEs directly and solely from an analysis point of view, we discuss which quantities should be estimated in the context of safety data, leading to the concept of safety estimands. Although the current debate on estimands and their role in clinical trials appears to be mainly related to efficacy evaluation \citep{Akacha2015,Leuchs2015}, the concept is applicable to safety endpoints as well. Only very recently this perspective has found its way into publications such as \citet{Akacha2017a}. Within the framework of estimands, we present estimation functions for estimating the quantities of interest, that is, statistical methods that map the AE data to a single value. In this context, we also describe problems related to meta-analyses of AE data and sketch possible solutions. Finally, we provide a review of current practices of health technology assessment (HTA) agencies with respect to the evaluation of safety data.\newline
An AE is any unfavourable and unintended sign including an abnormal laboratory finding, symptom, or disease temporarily associated with the exposure to an investigational product, whether or not considered related to the product \citep{ICH1995a}. The term ``treatment emergent'' is often added to an AE as a modifier in order to remove manifestations of preexisting conditions from consideration \citep{ICH1998}. Adverse events are documented by the investigator and coded with the Medical Dictionary for Regulatory Activities (MedDRA), which provides clinically validated medical terminology (https://www.meddra.org/). The MedDRA includes symptoms, diseases, diagnoses, investigation names and qualitative results, medical and surgical procedures, social and family history. Adverse events are coded with the ``lowest level terms''. These are combined for the analyses to so-called ``preferred terms''. The latest MedDRA version 21.0 contains more than 22000 preferred terms. As the present paper focuses on methods for analyzing AEs, we do not report further on definitions related to AEs or standards for the collection and documentation of AE data.\newline
The remainder of the paper is organized as follows. To begin with, our work is motivated in Section 2 by giving a brief overview of examples with respect to the analysis of AE data in clinical trials in different indications. 
 We also provide a review of current practices of HTA agencies with respect to the evaluation of AE data. The estimand framework is established in Section 3. We discuss what should be estimated in the context of safety data both from a regulatory context and from an HTA perspective. In Section 4, we provide statistical methods for analysing AEs with the focus being on the time to the occurrence of first AEs. We also describe some problems and their solutions for meta-analyses of AE data. In Section 5, we state some recommendations which estimators fit best to the described estimands. A discussion in Section 6 concludes the paper.

\section{Current practice of regulatory and HTA agencies}

\subsection{Examples}
Based on the experience with early benefit assessments by the Institute for Quality and Efficiency in Health Care (IQWiG) in Germany we looked over some examples whether variable follow-up times for AEs between individual patients, treatment groups and studies are common across different indications.\newline 
In general, trials with a primary time-to-event endpoint usually have variable follow-up times for each individual patient. However, depending on the indication the average follow-up time between treatment groups can be very different, see Table 1.
\begin{center}
*** Insert Table 1 about here ***
\end{center}
In oncology, study treatment is often given until disease progression only with limited follow-up time for AEs after discontinuation of trial treatment due to subsequent therapies, resulting in differential follow-up times for the different treatment arms and censored observations for the occurrence of AEs. Due to the dependency of follow-up time of AEs on progression-free survival, the treatment group with longer progression-free survival has a higher likelihood of observing an AE.
In such situations, a simple comparison of incidence proportions between arms is biased in favour of the inferior treatment. In time-to-event trials with average follow-up times not considerably different between treatment groups, there can still be variable follow-up times for individual patients. Examples are large trials in cardiovascular disease, metabolism (diabetes) or respiratory disease (COPD) with cardiovascular outcomes or mortality as primary endpoint, where mortality is relatively low.\\
Other trial designs than time-to-event trials with variable follow-up times were identified in infectious diseases and central nervous system disorders. For example, there are several trials with planned trial length in Hepatitis C that allowed shortening the treatment time either for all patients in the experimental arm or for those in the experimental arm who achieved an early response. Therefore, the follow-up time in the arm with the experimental drug was shorter than in the control group, see Table 1. Here, the follow-up times not only differ between the treatment groups but also between studies of a drug in the same indication (e.g. Hepatitis C). Comparison of AEs between treatment groups can be undertaken via incidence proportions only for the duration of the shorter treatment. It is not possible to demonstrate a potential advantage in terms of lower AE probabilities over a longer period of time. On the other hand, safety evaluations using na\"{\i}vely all AEs on treatment are biased in favour of the treatment group with shorter treatment duration. For instance, in multiple sclerosis, trials with fixed treatment duration have frequently been used, but if control patients switch to treatment in an extension study this would lead to longer follow-up in the experimental group and AE probabilities that are biased in favour of the control group. Figure 1 illustrates different scenarios of typical AE follow-up periods in clinical trials.
\begin{center}
*** Insert Figure 1 about here ***
\end{center}
Adverse events are assessed on a regular basis at the visit at the beginning of the trial (V0) and during treatment (V1,...,Vn). The end of treatment (EoT) triggers a safety follow-up visit (Saf-FU) marking the last regular safety assessment. Adverse events occurring on treatment or during safety follow-up are analyzed as treatment-emergent AEs (TEAEs, marked by bold symbols). First occurrences of AEs (marked by triangles) are in general considered only when occurring during the TEAE period. Serious AEs may be reported spontaneously after the safety follow-up visit.

\subsection{Regulatory context}
Drug approval usually requires a confirmatory proof of efficacy in at least two well conducted randomized controlled trials followed by a benefit-risk assessment that shows that the treatment's benefit outweighs the expected risk associated with the new treatment. In contrast to the demonstration of efficacy as compared to a control, the benefit risk assessment is far less standardized with respect to properly balancing benefit and expected side effects. Nevertheless, details on clinical trial specifications with respect to the investigated population, the study duration and the number of patients to be studied are given in several regulatory guidelines, as in therapeutic EMA guidelines (the \textit{CHMP clinical efficacy and safety guidelines}), ICH guidelines \citep{ICH1995c,ICH1995b,ICH1998c,ICH2016a} and U.S. FDA guidelines \citep{FDA2005a}.\newline
Safety assessments in drug approval are usually done with respect to the number and proportion of patients with specific AEs that occurred in the individual clinical trials, primarily focusing on the estimated probability of experiencing a given event within the predefined study duration potentially stratified in relevant subpopulations. It is based on a limited database available at the marketing authorization application where uncertainties about important risks may either prevent from authorization or imply the requirement of post-authorization safety investigations. Due to a number of different side effects, and since the absence of evidence of an increased risk is not necessarily evidence of absence of an increased risk \citep{altbland1995}, the analysis of AEs is often crude and merely descriptive aiming to detect safety signals, although proposals have been made in the literature on the use of more sophisticated methods \citep{allignol2016adv,bender2016biometrical,Chuang_Stein_2001}. Study duration may be too short or different studies may have different durations leading to different rates. In addition the database from phase III studies may be insufficient to detect infrequent but serious events. In case of uncertainty the EMA's Pharmacovigilance Risk Assessment Committee (PRAC) may require that a post-authorisation safety study (PASS) be carried out after a medicine has been approved. Nevertheless, difficulties in the assessment at the time of marketing authorization are highly relevant especially in indications with a small patient population and a high unmet medical need imposing high pressure to the regulatory system.\newline
Hence, an assessment of safety often targets the probability of an AE within a subject and a given period of time. The safety of a new treatment is assessed using all data available for this treatment. The FDA, e.g., asks for a document called integrated summary of safety (ISS) as described in \citet{FDA2009a}. In contrast to efficacy assessments, comparative safety assessments are difficult and the appraisal of side effects is usually done in absolute terms. Even if relative (comparative) safety assessments are given and relate to individual studies, accounting for different observational times due to study discontinuation, other events and the presence of changing individual proneness over time is difficult. Competing risks for a number of targeted adverse effects add another layer of complexity. 
For example, in two trials involving patients with type 2 diabetes and an elevated risk of cardiovascular disease, patients treated with canagliflozin had a lower risk of cardiovascular
events than those who received placebo but a greater risk of amputation of toes, feet, or legs with canagliflozin than with placebo (6.3 vs. 3.4 participants with amputation per 1000 patient-years, corresponding to a hazard ratio (HR) of 1.97; 95\% confidence interval: [1.41, 2.75]) \citep{Neal2017a}. In such a situation, the increased risk of amputation might be difficult to determine because of the much larger risk of mortality in comparison. 

\subsection{HTAs}
\subsubsection{The international perspective}
The evaluation of safety is considered as an important element of health technology assessments (HTA) \citep{Busse2012a}. However, due to different approaches to HTA in different health care systems and little methodological guidance given generally \citep{Pieper2014}, the integration of safety data and with it the analyses and interpretation of safety data differs considerably.\newline
Some health care systems, such as the one in England, focus on the economic value of a new technology by implementing a cost effectiveness threshold. Those are usually based on an incremental cost effectiveness ratio (ICER) or quality-adjusted life years (QALY) by comparing against the standard of care \citep{nicemanual}.\newline
In this context data on AEs which are considered most relevant from the patients' QoL and/or costing perspective are usually integrated by means of utility functions \citep{Ara2012a}. 
A comprehensive review of the current practice of economic models concludes that there appears to be an implicit assumption within modelling guidance that adverse effects are very important. There appears to be a lack of clarity how they should be dealt with and considered for modelling purposes \citep{Craig2009a}.\newline
Other health care systems, such as the one in Germany, base their decision on the incremental medical benefit against the standard of care \citep{IQWIG2017}. Usually data from clinical trials are used in order to evaluate the added benefit for all patient relevant endpoints separately to demonstrate an overall added benefit of the drug for the population in scope. Typically, a comprehensive description of AEs is provided in those assessments. However, no specific guidelines with respect to safety analyses are in place in countries following this approach. In the following, we discuss one specific example.

\subsubsection{The current practice at IQWiG in Germany}
At the beginning of 2011 the early benefit assessment of new drugs was introduced in Germany with the Act on the Reform of the Market for Medicinal Products (AMNOG). The Federal Joint Committee (G-BA) generally commissions the Institute for Quality and Efficiency in Health Care (IQWiG) with this type of assessment, which examines whether a new therapy shows an added benefit (a positive patient-relevant treatment effect) over the current standard therapy. The IQWiG is required to assess the extent of added benefit on the basis of a dossier submitted by the pharmaceutical company responsible. In this assessment, the qualitative and quantitative certainty of results within the evidence presented, as well as the size of observed effects and their consistency, are appraised. The general methods of IQWiG are described in the Allgemeine Methoden \citep[Version 5.0]{IQWIG2017}. In accordance with § 35b (1) Sentence 4 Sozialgesetzbuch (SGB) V, the following outcomes related to patient benefit are to be given appropriate consideration: increase in life expectancy, improvement in health status and quality of life, as well as reduction in disease duration and adverse effects. In the benefit assessment, all patient-relevant endpoints play a role, and for safety, particular consideration is given to serious and severe AEs as well as treatment discontinuations. In addition, AEs that are of special interest within the context of the disease or drug class considered may play a role.\newline
For the assessment of the extent of added benefit the effect sizes are of main interest. An effect size in this context is defined as the (relative) difference between the new treatment and the appropriate comparator therapy. It is an important step to grade the qualitative certainty of the estimated effect size e.g. based on trial design, data quality and estimation method. 
Patients not included into the analyses and patients with incomplete data increase the risk of bias. Therefore, the following information is gathered and considered to assess the risk of bias: study design (randomized, open-label or double-blind), proportion of patients without consideration in the analyses per study arm, proportion of patients with incomplete data per study arm (censored data, lost-to-follow-up, ...), reasons for censoring (informative, non-informative, competing risks), distribution of censoring times. Generally, this information is used to assess the direction (in favour of arm $x$) and the strength (low or high) of the risk of bias.\newline
In case of varying follow-up times, 
methods based on survival time analyses are preferred compared to analyses based on four-fold contingency tables \citep{bender2016biometrical}. To classify the risk of bias the number of the censored patients and the reasons for censoring have to be considered. Different types of censoring are possible: ``uninformatively censored'', ``patients with competing risks'' and ``informatively censored'', which influence the risk of bias. We will elaborate further on this issue in Subsection 4.3. 

\section{Estimands}
\subsection{Framework}\label{subsec:4.1:estima:frame}
It is paramount to agree upon the relevant target of estimation defined by the question what would happen to a specific patient or what is the patient's risk with respect to a specific event or multiple events when treated with a given drug as compared to another drug or to not being treated at all. In the context of \textit{efficacy} assessments, this concept has recently been introduced within the framework of \textit{estimands} in the new draft addendum R1 to the ICH E9 guideline on statistical principles in clinical trials entitled \textit{Estimands and Sensitivity Analyses in Clinical Trials}; this addendum is referring to the precise parameter or function of parameters to be estimated in situations where \textit{intercurrent events} as treatment discontinuation, death, rescue medication or switch to the other study treatment may influence subsequent measurements \citep{ICH2017,ICH1998}. We would like to stress the fact that the above-mentioned addendum is not yet finalized, hence our expositions in the sequel can only reflect the current state of discussion. Parts of the draft addendum are seen critically by HTA agencies \citep{IQWIG2018}, and the addendum does not prescribe the use of a particular estimand in a certain situation. Moreover, the discussion of the application of the estimands approach to safety data and questions related to benefit-risk assessments is ongoing and only started recently.\newline
Four different elements are required to describe the estimand of interest: the \textit{targeted population}, the \textit{endpoint} (variable), the \textit{intervention effect} that describes how intercurrent events that potentially influences the endpoint are accounted for, and the \textit{summary measure} that summarize the comparison of the two treatments under investigation.\newline
Whereas the nomenclature of types of estimands has been developed and changed during the last few years, currently, the following classes of estimands are discussed in the regulatory context:
\begin{itemize}
\item
{\bf Treatment policy}: \textit{treatment policy estimand} do not account for any intercurrent event. The treatment effect is measured irrespective of any intercurrent event, as treatment discontinuation or additional medication given. 
\item
{\bf Composite}: \textit{composite estimands} combine the variable of interest with the intercurrent event, e.g., by defining a treatment failure by the lack of response or treatment discontinuation.
\item
{\bf Hypothetical}: \textit{hypothetical estimands} target an effect that would occur in the overall population in a hypothetical scenario, in which no patient experienced the intercurrent event. E.g., the effect of all patients adhering to treatment constitutes a hypothetical effect when some patients in fact do not adhere to treatment. 
\item
{\bf Prinicpal stratum}: \textit{principal strata estimands} are defined by the subset of patients in whom the intercurrent event occurs either under one of the treatments or under both. Since a group comparison trial cannot directly identify these patients with respect to the not-administered treatment, causal inference methods using specific assumptions would be required for the analysis.
\item
{\bf While on treatment}: \textit{while on treatment estimands} relate to the effect prior to the occurrence of an intercurrent event, e.g. before intake of rescue medication or the effect while being alive.
\end{itemize}
Although the current discussion is related to the \textit{efficacy} evaluation, the concept is applicable to \textit{safety} endpoints, usually the occurrence of a specific side effect, as well. Considering the variable of interest as the time to a specific side effect, summary measures might be given after a specific period of time. Relevant \textit{intercurrent events} are treatment discontinuation or switch, death or other side effects that may prevent from the event of interest.\newline
Whereas the basic idea of an estimand is not restricted to the efficacy assessment, different issues related to the large number of event types and the desired ``equivalence proof'' combined with the cautionary principle point to somewhat different difficulties. Envisaging the chances of a beneficial treatment in the sense of both, efficacy and safety, for a given patient may be seen as a concept of incorporating efficacy and safety in an estimand, but may be still difficult to be interpreted in the comparison of two treatments with different safety and efficacy patterns. In that sense, it appears sensible to go along the lines that are conceived for the efficacy assessment and clarify the precise parameters in the event analysis in the presence of other concurring events either in relation to the given treatment, the patient's condition or competing side effects. 

\renewcommand{\baselinestretch}{1.0}
\subsection{Estimands in the regulatory context and in HTAs - what to aim for?}
\renewcommand{\baselinestretch}{1.5}
The estimand framework is not specific to the clinical development of new drugs in the regulatory context but is also relevant to address needs of HTA bodies. The aim of the HTA process is the assessment of evidence as a basis for further decisions about reimbursement, pricing and market access.\newline
Estimands are supposed to focus and describe the research question in detail. As the aims of drug approval agencies and HTA bodies differ to some extent, different estimands may be of primary interest, but some overlap in secondary considerations can be expected. The following considerations focussing on HTA may also apply to the benefit risk assessment in drug approval. In the German early benefit assessment according to AMNOG, i.e. §~35a SGB V, there is a need for the HTA authorities to identify the estimands, which are not necessarily the same as in the regulatory context to obtain marketing authorization.\newline
For marketing authorization the current practice, in general, is to report estimates for what could be considered while on treatment estimands to provide evidence for the safety profile of the treatment of interest. Specific information, however, on AEs occurring in a long-term follow-up regardless of study drug adherence may be requested in special cases, which may, however, be hampered by the limited observational period. Potentially diluting effects of treatment policy estimands in case of treatment discontinuation or switch may, depending on the treatment comparison and the disease, be anti-conservative for comparative safety assessments, hence favouring while on treatment estimands for regulatory purposes. Certainly, within the context of recent regulatory discussion on estimands in efficacy, a new regulatory framework also on estimands in safety is needed.\newline
HTA bodies are most interested in the \textit{treatment policy estimand}, independently of occurrences like rescue therapy (e.g. rheumatoid arthritis) or subsequent therapies (e.g. oncology). However, in indications like oncology, the AEs are frequently only collected up to a certain point after last dose of study medication. These data do not support a treatment policy estimand. In such situations, the evidence for risk assessment is less strong. It may be difficult or impossible to cover the treatment policy estimand with the data usually obtained. To obtain sufficient data for a treatment policy estimand and provide a solid basis for an early benefit assessment, the current practice of how data are collected in clinical studies needs to be changed \citep{bender2016biometrical}.\newline
A major challenge in oncology is the treatment change after progression of the disease, where in many cases patients enter a subsequent clinical study, e.g. in malignant melanoma where about four years ago the only treatment option was dacarbacin and most patients entered clinical trials after progression. These studies were under evaluation by the HTA bodies recently due to the time gap between study conduct and marketing authorization. However, in most studies a patient is not allowed to enter a new clinical study, if they are still participants of the prior study. As a consequence, they need to withdraw consent for the first study to enter the next. Therefore, AEs cannot be collected after progression for the first study in general. Exceptions from this practice are observed \citep{Robert2015}, following a protocol recording new onset of serious adverse events up to 90 days after last dose of study treatment and those serious AEs considered related any time after discontinuation of treatment. Another example consists of the German Society of Pediatric Oncology and Hematology (GPOH) \citep{Calaminus2007}, which records further follow-up data on children after end of treatment and with this provides the possibility for long-term surveillance and follow-up and late effect evaluation in paediatric oncology patients. Apart from these examples, the while on treatment estimand is used in clinical trials to avoid bias due to unbalanced withdrawals in the treatment groups. E.g., if the control treatment is a standard first line treatment and a subsequent study in second line requires the standard treatment as first line, then only patients of the control group are allowed to enter the second line study leading to unbalanced subsequent therapy along with biases in efficacy but also safety.\newline
Four different scenarios, which are displayed in Figure 2, can be distinguished to describe safety estimands in an HTA system.
\begin{center}
*** Insert Figure 2 about here ***
\end{center}
The scenarios in Figure 2 differ according to the lengths of the planned and observed follow-up times in the study. The definition of estimands becomes increasingly complex with more pronounced differences in follow-up time due to intercurrent events. In the first two scenarios, the planned follow-up times of AEs in the study population are similar, e.g. in trials with fixed trial lengths (see Subsection 2.1). Whereas in Scenario 1, there are no or only minor differences in observed follow-up times between individual patients within treatment groups or between treatment groups, in Scenario 2 medium to large differences in observed follow-up times do exist, due to e.g. high level of treatment or study discontinuations. Scenarios 3 and 4 consider studies with expected differences in follow-up times by treatment group, e.g. in oncology (see Subsection 2.1), where AE reporting stops a certain number of days after last dose of study drug and the time on study drug differs between treatment groups. In all four scenarios, HTA bodies usually aim for the treatment policy estimand. However, studies are commonly planned to collect data that are appropriate for the while on treatment estimand, which is usually the focus of regulatory agencies in the marketing authorization process. When benefit dossiers are based on the same studies, it is often not possible to provide estimates of the treatment policy estimand desired by the HTA agencies due to lack of adequate data. Hence, it is apparent that the described requirements by HTA bodies with regard to safety analyses need to be taken into account already in the planning phase of a clinical study.

\section{Statistical methodologies}\label{sec:5:statmeth}
The aim of this Section is to discuss methods for \textit{analysing} AEs. We focus on the occurrence of the first AE of a specific type because the statistical considerations for the analysis of the first AE are relevant also for the analysis of recurrent AEs. We found a great variety of methods in the literature, some of which were not well defined, making it difficult to
identify both estimator and estimand. Main methods which focussed on adverse
events occurrence in one group were described in the literature as crude rate,
incidence proportion, incidence rate, exposure-adjusted incidence rate, hazard
function for adverse events, Kaplan-Meier and cumulative incidence function
considering competing risks. We first describe methods of estimation within
one treatment group in Subsection~\ref{subsec:one}, and subsequently the
comparison of AE occurrence between samples in Subsection~\ref{subsec:two}. The time point of AE occurrence or
comparison thereof will be made explicit. A major issue will be that safety
data need not be completely observed over the whole study period for all patients and will possibly be right-censored. This requires
the use of time-to-event methodology \citep{Chuang_Stein_2001,allignol2016adv} taking different follow-up times into account,
which is common in efficacy analyses, but less so for safety. In
  this context, it is often discussed which kind of censoring is informative,
  and the role of censoring will briefly be revisited in Subsection~\ref{subsec:three}. Methods for meta-analyses of AE data are discussed in Subsection 4.4.

\subsection{Methods of estimation within one treatment group}\label{subsec:one}

One major common method is the so-called ``crude rate'', which despite its name is in fact a proportion and is defined as
$\hat{\p}(\adv)= a/n$, where $a$ is the number of patients observed to experience at least one AE of a
  specific type and $n$ is the total number of study patients, see \cite{allignol2016adv},
\cite{Chuang_Stein_2001}, \cite{Siddiqui_2009},
\cite{Ioannidis_2004}, \cite{Lineberry_2016} and references therein. The crude rate is a correct estimator of
the probability to experience at least one AE of the interesting
  type in case of complete data and identical follow-up times for
  all patients.
With different follow-up times in different samples,
the crude rate will estimate the AE probability at different points in
time. This difficulty is resolved by considering the incidence proportion (\cite{allignol2016adv} and references therein):
 \begin{equation} \label{pint}
   \hat{\p}(\text{AE in $[0, t]$}) = \frac{\sum\limits_{u \le t} a_u}{n},
\end{equation}
where $a_u$ denotes the number of patients observed to experience at least one AE of a specific type at time $u$.
The expression (\ref{pint}) estimates the probability of experiencing at least one AE within some
time-interval $[0,t]$, but is again only valid for complete data over the considered time interval. In the
presence of censoring, both the crude rate and the incidence proportion
\emph{underestimate} the AE probability \citep{allignol2016adv}. The reason is
that these methods in fact estimate the probability of
\emph{both} the AE occurrence
\emph{and} the non-occurrence of censoring.\newline
Some authors, e.g.  \citet{Ioannidis_2004} and \citet{Amit_2008}, therefore
suggest to use one minus the Kaplan-Meier estimator for estimating
$\p(\text{AE in $[0, t]$})$, censoring time to AE by both the end of follow-up and by competing events that preclude AE
occurrence such as death without a prior AE. However, this method
\emph{overestimates} the AE probability
\citep{allignol2016adv,gooleyEstimation1999}. The reason is that one minus the Kaplan-Meier estimator approximates a cumulative distribution function, implicitly
assuming that eventually 100\% of all patients experience the AE under
consideration, possibly after death. Therefore, one minus the Kaplan-Meier estimator must not be used to estimate $\p(\text{AE in $[0, t]$})$ in the presence of competing events which prevent the occurrence of the AE under consideration.\newline
It is the Aalen-Johansen estimator \citep{Aale:Joha:an:1978}
that generalizes the Kaplan-Meier estimator to multiple event
  types and nonparametrically estimates the so-called cumulative incidence
function $\p(\text{AE in $[0, t]$})$, accounting for both competing events \citep{oneill1987,Crowe_2013} and the usual censoring
due to end of follow-up. The Aalen-Johansen estimator for the probability of AE occurrence is (\cite{allignol2016adv} and references therein):
\begin{equation} \label{ajest}
  \hat{\p}(T \leq t, \adv) = \sum_{u\le t} \hat{\p}(T > u-)\cdot\frac{a_u}
  {n_u} \enspace ,
\end{equation}
where $T$ is the time until occurrence of an AE or of a competing event, $\hat{\p}(T > u-)$ denotes the estimate of the probability of not experiencing an AE or the competing event just prior to time $u$ and $n_u$ is the number of patients at risk of observing an AE or a competing event just prior to $u$. The interpretation of (\ref{ajest}) is that of a sum over the empirical
probabilities of experiencing an AE at the observed event times. Here,
for estimation of ${\p}(T > u-)$ the Kaplan-Meier
  method is used, because the definition of~$T$ encompasses all
  competing events.\newline
Time-to-event analyses are based on hazards, because in general follow-up times are incomplete. In
fact, the sum over the quotients on the right hand side in (\ref{ajest})
is the Nelson-Aalen estimator of the cumulative AE hazard $\int_0^t
\alpha_{\adv}(u)du$. For analysing AEs, the Nelson-Aalen estimator is key in
three ways. Firstly, it enters the computation of the Aalen-Johansen
estimator. Secondly, it is closely linked to the Mean Cumulative Function
which is based on the Nelson-Aalen estimator and is also used in safety
analyses \citep{Siddiqui_2009}. Thirdly, the Nelson-Aalen estimator is the
cumulative nonparametric counterpart of the commonly used incidence rate (or
incidence density) of AEs \citep{allignol2016adv,
  Ioannidis_2004,Lineberry_2016}:
\begin{equation} \label{irae}
IR_{\adv} = \frac{a}{\sum t_i}
\end{equation}
where $t_i$ is the time at risk for patient $i$ and $\sum t_i$ denotes the population time (person-years) at risk. The incidence rate (\ref{irae}) is an estimator of the AE hazard $\alpha_{\adv}(t)$ under a constant
hazard assumption, $\alpha_{\adv}(t) = \alpha_{\adv}$ for all times $t$. The
incidence rate is popular, because its denominator accounts for varying
follow-up times. Sometimes, the exposure-adjusted incidence rate is reported by counting in the denominator only the population time during exposure to study treatment. However, it is \emph{not} a
probability estimator (and should better not be reported as a percentage). In
fact, it is easily seen that depending on how time is measured (think of
milliseconds or decades), the denominator can be made arbitrarily large or
small, possibly resulting in values larger than one.\newline
In perfect analogy to the Aalen-Johansen estimator, translating incidence
rates into probability statements requires incorporating competing events (CEs), e.g. death without prior AE. For
instance, $IR_{\scriptsize{\mbox{AE}}}$ and the incidence rate of the competing event, $IR_{\scriptsize{\mbox{CE}}} = c / \sum t_i$ (writing $c$ for the number of patients observed to experience the competing event), can be used to obtain a parametric counterpart of the Aalen-Johansen
estimator. If constant event-specific hazards are assumed, the cumulative incidence function of the event type AE is explicitly given as
\begin{eqnarray}\label{cifpara}
 \mbox{P}(T \le t, \mbox{AE}) & = & \int_0^t \alpha_{\adv} \cdot \exp \left( - ( \alpha_{\adv} +  \alpha_{\scriptsize{\mbox{CE}}} ) s \right )  \: \mbox{d}s \\ \nonumber
 & = & \frac{\alpha_{\adv}}{ \alpha_{\adv} + \alpha_{\scriptsize{\mbox{CE}}}} \left(1-\exp(-(\alpha_{\adv} + \alpha_{\scriptsize{\mbox{CE}}})t) \right ) \enspace ,
\end{eqnarray}
where $\alpha_{\scriptsize{\mbox{CE}}}$ denotes the competing event hazard.
By plugging in both event-specific incidence rates, the cumulative incidence function can be estimated \textit{parametrically} under a constant hazards assumption.\newline
We also note that both the Nelson-Aalen estimator and the incidence rate allow
for AEs to be recurrent. In this situation, the translation into probability statements becomes
more complex because of the more complicated recurrent events structure, but also with recurrent AEs competing events have to be taken into account.

\subsection{Comparison of treatment groups}\label{subsec:two}

When comparing two treatment groups with respect to AE occurrence, often
measures like risk difference, relative risk or odds ratio of crude rates are suggested \citep[e.g.][]{Amit_2008}. However, if such relative measures are used in the
presence of censoring and are based on biased one-sample estimators as
discussed above, the result of such a comparison will be biased too, but the direction of the bias is uncertain. For instance,
a ratio of incidence proportions calculated from censored data will divide
something too small by something too small.\newline
As a parametric analysis, the ratio of incidence rates is an appropriate estimator of the hazard ratio under a constant hazard
assumption. The obvious semi-parametric extension is to use a Cox proportional hazards model,
\begin{equation}
  \label{eq:cox_ae}
 \alpha_{\adv}(t|Z) = \alpha_{\adv;0}(t)\exp(\beta_{\adv}^{\top} Z) \enspace , 
\end{equation}
where $\alpha_{\adv;0}(t)$ is an unspecified baseline AE hazard,
$\beta_{\adv}$ is the vector of regression coefficients and $Z$ a vector of
baseline covariates including treatment group. In other words, if in (\ref{eq:cox_ae})
  the only covariate is treatment group, $Z\in \{0,1\}$, then the
  ratio of the incidence rate in group 1 to the incidence rate in group 0
  estimates the hazard ratio $\exp(\beta_{\adv})$ under the
  assumption of a constant baseline hazard for adverse events,
  $\alpha_{\adv;0}(t)\equiv$ constant. If this assumption is in
  doubt, any Cox regression software \emph{technically} censoring the time to
  AE by observed competing events will yield the usual maximum partial
  likelihood estimator of $\exp(\beta_{\adv})$. Technically, censoring by observed competing events is in perfect analogy to
  calculation of the incidence rates, but, again in analogy to the
  incidence rates, it does not allow for probability statements. In other
  words, the analysis remains somewhat incomplete without consideration of the
  hazard of the competing event, e.g., via a second Cox model,
  \begin{displaymath}
  \alpha_{\scriptsize{\mbox{CE}}}(t|Z) = \alpha_{\scriptsize{\mbox{CE}};0}(t)\exp(\beta_{\scriptsize{\mbox{CE}}}^{\top} Z) \enspace ,
  \end{displaymath}
  which \emph{technically} censors the time to the competing event by observed AEs, see
\citet{BAM} for a practical in-depth discussion. A reasonable method of choice will be a Cox regression model for the event-specific hazards. The important point is that it requires as many Cox regression models as there are event-specific hazards present. Although fitting two Cox models is straightforward from a
  computational perspective, the presence of two hazards is not without
  subtleties.\newline
  We want to illustrate this using a toy example, assuming, for
    ease of presentation, constant hazards.  We consider a treatment that modifies the AE hazard by a factor of $0.5$ and
    the competing event hazard by a factor of $0.25$.  As $t \rightarrow \infty$, one can see from (\ref{cifpara}) that
  $\p(\adv\ |\ \text{group 1})$ in treatment group~$1$
becomes
\begin{equation} \label{padv1}
\frac{0.5 \cdot \alpha_{\adv_0}}{0.5 \cdot \alpha_{\adv_0}+0.25 \cdot \alpha_{\scriptsize{\mbox{CE}}_0}} \enspace ,
\end{equation}
where $\alpha_{\adv_0}$ and $\alpha_{\scriptsize{\mbox{CE}}_0}$ denote the AE hazard and competing event hazard in group 0, respectively. The expression (\ref{padv1})
is greater than $\p(\adv\ |\ \text{group 0})$,
although the AE hazard has been reduced. The reason is simple. In our toy example, treatment reduces both hazards, thus, delaying both events. Because
the effect is larger on the competing event than on the AE, there will eventually be more AEs in treatment group 1 than in treatment group 0,
such that the cumulative AE probabilities cross at some point in time. This is illustrated in Figure 3, showing the cumulative AE probabilities in group 0 and group 1 over time for the situation of constant hazards described above.
\begin{center}
*** Insert Figure 3 about here ***
\end{center}
In group 0, both the AE hazard rate and the competing event hazard rate were set to 0.02 events per day, eventually leading to an AE probability of 1/2 in group 0 and of 2/3 in group 1. In group 1, the AE hazard is modified by a factor of 0.5 and the competing event hazard by a factor of 0.25.\newline
Because multiple hazards are present and an analysis of only one hazard does
not suffice for probability statements, so-called `direct' approaches
such as the Fine and Gray model for the so-called subdistribution hazard \citep{Fine:Gray:prop:1999} or,
easier to interpret, the proportional odds cumulative incidence function model
\citep{no_fine_gray_2015} have been developed. Most popular is
perhaps the Fine and Gray approach, which interprets one minus the cumulative
incidence function as a survival function and fits a Cox model to the
corresponding hazard, the so-called subdistribution hazard. The approach is
useful in that a subdistribution hazard ratio greater (smaller) than one
translates into an increase (decrease) of the cumulative incidence function
but is otherwise difficult to interpret \citep{andersen2011interpretability},
because the subdistribution hazard, say~$\lambda(t|Z)$, can be expressed as
\begin{displaymath}
  \lambda(t|Z) = \frac{\p(T>t|Z)}{1-\p(T \leq t, \adv|Z)} \cdot \alpha_{\adv}(t|Z),
\end{displaymath}
which results in a complicated mixture of effects on the hazard scale and on
probability scales. Alternatives include group comparisons based on confidence
bands of the cumulative incidence functions \citep[e.g.][]{beyersmann2013weak} or the proportional odds cumulative incidence function
model \citep{no_fine_gray_2015} mentioned above. The latter is a generalization of the logistic
regression model to binomial probabilities $\p(T \leq t, \adv|Z)$ as a function of
time~$t$ and in the presence of censoring.

\subsection{Censoring: independent or informative?}\label{subsec:three}
Our presentation so far has demonstrated that the analysis of AE occurrence in
trials with varying follow-up times has to account for differences in
follow-up, in particular in the form of \textit{censoring}. Survival methodology should
not only be used for efficacy, but also for safety analyses. However, the safety
estimands of interest may still be a matter of debate (see Section 5). Above, we have
demonstrated that the relationship between hazards and probabilities is more
subtle when competing events that preclude AE occurrence are present. But even when this is
accounted for, there may be a choice between, say incidence rates and
`exposure-adjusted incidence rates' (which are also incidence rates, but with
a different at-risk period, see Subsection 4.1).\newline
Censoring is a concept that pertains to all these aspects, but it
  is more subtle than may seem at first glance. So far, we have argued that
  more standard statistical techniques not from the field of survival analysis
  are inappropriate because the analysis will be about AEs that are
  \emph{observed} rather than about AEs that the patient experiences. The observation time of AEs is often restricted, e.g. in oncological trials often to progression of disease + 30 days. The patient may experience AEs after this period but this is not reported in the case report form (CRF). Next, we
  have found that one minus the Kaplan-Meier estimator censoring
  the time to AE by competing events overestimates the
  cumulative AE \emph{probability}, but that such a censoring approach yields
  a valid analysis of the AE \emph{hazard}. Whether or not
    censoring yields a valid analysis, has an impact on the estimand at
    hand, of course.\newline
  When survival methods for AE analysis are discussed, authors often warn
  against \emph{informative censoring} \citep[e.g.][]{bender2016biometrical}, but this discussion on censoring is somewhat
  complicated by inconsistent terminology in the literature. \emph{Random
    censoring} typically refers to the situation where time-to-event and
  time-to-censoring are independent random variables taking values
  in~$[0,\infty)$ \citep[e.g.][p. 30]{ABG}. This is also called
  \emph{independent censoring} or \emph{non-informative censoring}
  \citep[e.g.][]{o2008proportional} in the literature and it is neither
  uncommon that these last two terms are used interchangeably \citep[e.g.][]{Collett2015a} nor that they refer to different censoring
  mechanisms \citep[e.g.][]{kleinbaum2010survival}. This bedevils the
  discussion both on AE analyses and estimands, because one must first define
  what is meant by, e.g., the term \emph{independent censoring}, because
  different authors may use the term differently, referring to different
  censoring mechanisms.\newline
  In our context, one will rarely be willing to assume that the time to a certain AE and the
  time to a competing event such as death (or progression) without prior AE
  are independent and present themselves as an example of random
  censoring. In fact, and more importantly, it is entirely unclear how to
  define time-to-AE for a patient who has died as the value of a random
  variable in the positive real numbers. Such a value would suggest that there
  is an AE \emph{after death}, which is an awkward concept (to say
  the least), and we prefer an agnostic point of view.\newline
  Above, we have found that observed occurrence of a competing event can be
  regarded as `independent censoring' in the sense that technically treating
  it as a censored observation allows for a correct analysis of the AE
  hazard. In other words, the analysis of the AE hazard \emph{does not depend}
  on whether censoring was due to administrative closure of the study or
  whether it was due to a competing event. On the other hand, observed
  occurrence of a competing event can be regarded as `informative censoring'
  in the sense that technically treating it as a censored observation does not
  allow for a correct analysis of the AE probability as in a Kaplan-Meier
  procedure.\newline
  These ideas are made rigorous in the counting process approach to survival
  analysis \citep{ABGK, andersen2006survival, ABG}, see also
  \citet{allignol2016adv} for a non-technical account. In a nutshell, censoring
  by a competing event is independent censoring in that it preserves the
  desired form of the intensity of the counting process of the event under
  consideration, but it becomes independent, yet informative censoring if the
  target parameter is the cumulative incidence function.\newline
  It is worthwhile to reflect on these concepts in oncology trials, where
  common endpoints are progression-free survival and overall survival. It is
  not uncommon that recording of AEs is stopped for patients who progress and
  undergo a second line treatment. Of course, these patients may still
  experience AEs, and it is generally assumed that the hazard of an AE after
  progression is different as compared to before progression. Progression then
  is a competing event for AE without prior progression. And progression is a
  competing event for death without prior progression \emph{and} without prior
  AE. Hence, censoring by the progression event will yield a
  valid analysis of the hazards of the other two competing events, but any
  probability statement will need to account for all hazards involved. A
  different question, however, is what kind of censoring by
  progression is with respect to AE occurrence \emph{after} progression. In a way, the
  answer is easy: if censoring by progression events yields a
  valid analysis of AE occurrence \emph{before} progression, but if the AE
  hazard \emph{after} progression changes, progression cannot be independent
  (and, hence, not non-informative) censoring with respect to AE occurrence \emph{after}
  progression.\newline
  The argument can be made rigorously by showing that censoring by
  progression does not preserve the desired form of the intensity of the AE
  counting process, if the latter is not restricted to AEs before progression,
  but the bottom line is obvious: if recording of AEs is stopped for patients
  with diagnosed progression, inference for post-progression AEs is
  impossible.

\subsection{Meta-analyses of adverse event data}
When data from more than one study are available it is not uncommon to na\"{\i}vely pool the data across the studies by e.g. ``simply combin[ing] the numerator events and the denominators for the selected studies'' \citep{FDA1996a}. \citet{McEntegart2000} and later \citet{Ruecker2008} as well as \citet{Chuang2011} warned of such na\"{\i}ve  pooling as results might be biased due to Simpson's paradox. The International Conference on Harmonization (ICH) E9 states that ``any statistical procedures used to combine data across trials should be described in detail'' and that ``attention should be paid [...] to the proper modelling of the various sources of variation''. The use of meta-analysis techniques is encouraged \citep{CIOMS10}, since these techniques allow for \textit{variation in baseline} (control group) outcomes across the various studies. Random-effects meta-analysis \textit{in addition} allows for \textit{variation in treatment effects} across studies (so-called \textit{between-trial heterogeneity}). Therefore, this type of models is appropriate to formally combine several studies in one analysis. In the context of safety analyses a number of specific problems arise \citep[see e.g.][]{Berlin2013}, some of which will be considered in the following.\newline 
Meta-analysis can be carried out using \textit{aggregated data} of the individual studies or, if available, \textit{individual patient data} (IPD). IPD meta-analyses have some advantages over aggregate data meta-analyses \citep{Tierney2015}, in particular with time-to-event data considered in this manuscript. If time-to-event data are considered and the meta-analysis is based on published data, it is sometimes necessary to reconstruct the data by using appropriate methods \citep{Guyot2012,Liu2015}.\newline
As explained in Subsection \ref{subsec:two}, effect measures such as the risk difference, relative risk or odds ratio of crude rates are not appropriate when analyzing AE with varying follow-up times. Alternatives include the ratio of incidence rates, which would be appropriate for instance under a constant hazard assumption, the hazard ratio estimated in a Cox proportional hazards model or the subdistribution hazard ratio of the popular Fine \& Gray model. For the purpose of meta-analyses these effect measures would be log-transformed to estimate a combined effect on the log-scale, e.g. log rate ratio or log hazard ratio. Technically this is fairly straightforward. However, some challenges arise if for example the follow-up times vary considerably between the studies. Under assumptions such as proportional hazards the formal combination of studies with different follow-up times are justified. In practice, however, such assumptions might be challenged. Whereas with a single study the hazard ratio estimated by weighted Cox regression might be interpreted as an average effect over follow-up when the proportional hazards assumption does not hold true \citep{Schemper2009}, this interpretation in the presence of considerably different follow-up times across studies in a meta-analysis does not apply in the same way. Furthermore, with these arguments the variation in follow-up times across studies is likely to yield some level of between-trial heterogeneity in treatment effects.\newline
Summarizing AE data from a clinical development programme, typically only a small number of studies is available. Meta-analysis of (very) few studies has recently attracted more attention as it is also frequently accounted in settings other than the one considered here. An overview and discussion of the various methods in the context of benefit assessments is provided by \cite{Bender2017}. Specifically, empirical studies demonstrated that between-trial heterogeneity is likely to be present \citep{Turner2012}, suggesting the use of random-effects rather than fixed effect meta-analysis. With \textit{few} studies, however, the \textit{between-trial heterogeneity} is difficult to assess with standard methods for random-effects meta-analysis based on normal approximation yielding confidence intervals for the combined effect which are too short and which have coverage probabilities well below the nominal confidence level. This is due to an underestimation of the between-trial heterogeneity and a failure to account for the uncertainty in the estimation of the heterogeneity. \textit{Bayesian} random-effects meta-analysis with weakly informative prior on the between-trial heterogeneity (and an uninformative prior on the treatment effect) has been suggested for meta-analysis with few studies \citep{Spiegelhalter2004,Higgins2009}. This avoids zero estimates of the between-trial heterogeneity and accounts for uncertainty in the estimation yielding satisfactory coverage probabilities and interval lengths \citep{Friede2017a,Friede2017b}. Application of the DIRECT algorithm \citep{roever2017}, which is faster than MCMC sampling and does not require inspection of convergence diagnostics, means that computations are fairly straightforward. An implementation is available in form of the R package \verb"bayesmeta", which can be downloaded from CRAN (\verb"https://cran.r-project.org/package=bayesmeta").\newline
\citet{Neal2017a} report an integrated analysis of two large-scale randomized placebo-trials assessing the efficacy and safety of canagliflozin in patients with type 2 diabetes and elevated risk of cardiovascular disease. Patients were followed up for varying lengths of time as the programme was event driven with a constraint on minimum follow-up. In a stratified Cox regression, a beneficial effect of canagliflozin versus placebo on the primary outcome time to death from cardiovascular causes, nonfatal myocardial infarction, or nonfatal stroke, whatever occurred first, was demonstrated (HR = 0.86; 95\% confidence interval (CI): [0.75, 0.97]). For the purpose of illustration, we consider here the AE ``low trauma fracture''. Figure 4 is a forest plot of the HR with 95\% CI from the two studies CANVAS and CANVAS-R.
\begin{center}
*** Insert Figure 4 about here ***
\end{center}
The Figure also includes a fixed-effect meta-analysis, which is in fact very similar to the stratified Cox regression reported by \citet{Neal2017a}. As noted by \citet{Neal2017a}, there was considerable between-trial heterogeneity. Therefore, the use of a random-effects meta-analysis is indicated. The forest plot includes results from three random-effects meta-analyses, modified Knapp-Hartung as a frequentist method suggested for meta-analyses with few studies \citep{Roever2015a} and Bayesian random-effects meta-analyses with two choices of the prior for the heterogeneity parameter $\tau$ \citep{Friede2017b}. As can be seen from Figure 4, the modified Knapp-Hartung method yields a very wide, non-informative interval whereas the Bayesian intervals are much shorter. In comparison to the fixed-effect model, the Bayesian intervals are considerably longer as they account for the pronounced between-trial heterogeneity.\newline
Meta-analyses of AE data are further complicated when the events considered are \textit{rare}. Normal approximations of the distributions of effects, e.g. log hazard ratios or log odds ratios, break down with low event rates. In particular if some studies result in zero events in both arms. Then measures such as (log-)odds ratios cannot be calculated. Among the remedies proposed for such problems are continuity corrections \citep{Bradburn2007} and models of the counts such as binomial distributions. The latter can be fitted using likelihood or Bayesian methods \citep{Spiegelhalter2004,Boehning2008,Kuss2015}. Very recently, the use of weakly informative priors for the treatment effect as well as for the between-trial heterogeneity has been shown to result in satisfactory properties in meta-analyses with few studies and rare events \citep{Gunhan2018a}.

\section{Estimators for described estimands}
In the following, the different statistical methods discussed in the previous
Section are put into the context of the estimand
framework described in Subsection~3.1 for the specific
situation of the analysis of AEs. The decision on the estimand is made on the study level. Although in the following we consider estimands of individual studies, the same principle applies to meta-analyses of AE data. A disadvantage of aggregate data meta-analyses in this context is of course that different estimands might have been used in the studies whereas in IPD meta-analyses the same or at least similar estimands can be applied to the studies. As with Section~3, our statements are based on the estimands as described in the draft addendum R1 to the ICH E9 guideline \citep{ICH2017}, hence the estimands we are referring to in this Section may be subject to change. \newline We focus specifically on trials in \textit{oncology}
with differences in follow-up times driven by progression, discontinuation of
treatment, death or end of follow-up. As progression in general leads to
treatment discontinuation, we consider only the intercurrent events \textit{treatment
discontinuation} and \textit{death}. For all estimands considered in
Subsection~3.1, except the principal stratum estimand,
the population is defined by the inclusion and exclusion criteria of the study
and contains the treated patients {\color{black}(first element of estimand
  description). With respect to the second element of estimand description,
  for all estimands, t}he endpoint is the time to the first AE of a specific
type. The estimands differ as a consequence of the following aspects: the
follow-up time over which AEs are included in the analysis and the way how the
intercurrent events {\color{black}treatment discontinuation and death are
  accounted for (third element of estimand description), and by the
  population-level summary for the endpoint (fourth element of estimand
  description).}  Here, by `intercurrent events' we really mean
`\textit{post-randomization events}', because death is both post-randomization and
terminal, but not truly intercurrent.\newline
Considering the \textit{treatment policy estimand}, the interest is in the comparison of
  treatment groups with respect to AE occurrence until death or end of
  follow-up, irrespective of the intercurrent event discontinuation of
  treatment. So, it includes all AEs until death or end of study, and, therefore, requires the
  collection of AE data after treatment discontinuation. The AE hazards of
the treatment groups can be compared by calculating the hazard ratio in a Cox
regression model where for patients without AE the time to AE is censored by
death or by end of follow-up. In the interpretation of the result, it has to
be considered if treatment has an effect on the competing event death without
prior AE, i.e. if the hazard ratio between treatment groups with respect to
death without prior AE is different from one. The hazards of the competing
event death without prior AE have also to be taken into account in the
estimation of the AE probabilities within the treatment groups by using the
Aalen-Johansen estimator. Treatment groups can also be compared with respect
to the AE probabilities by estimating the difference between AE probabilities
at a specified time point or over the whole study period, by calculating the
subdistribution hazard ratio from a Fine and Gray model
\citep{Fine:Gray:prop:1999}, or by calculating the odds ratio
from a proportional odds cumulative incidence function model
\citep{no_fine_gray_2015}. The decision between treatment
comparison by means of hazard functions or by means of probability functions
defines the fourth element of estimand description mentioned above.\newline
The \textit{while on treatment estimand} includes the AEs until discontinuation of treatment
and {\color{black} requires the collection of AE data up to this event. Treatment groups
can be compared with the same methods as used for the treatment policy estimand, now
treating discontinuation of treatment before AE and death without prior AE as competing events.}\newline
The \textit{composite estimand} would combine the interesting event AE with the
intercurrent events {\color{black}treatment discontinuation} and death without
prior AE. The endpoint is then time to AE, treatment discontinuation, or death
whatever occurs first. So, AE data after treatment discontinuation are not required. However, this does not hold in general for any intercurrent event. In this situation, no competing events are present and standard survival analysis
techniques can be applied, i.e. the composite event hazards of the treatment
groups can be compared by calculating the hazard ratio in a Cox regression model,
and the probabilities of the composite event can be estimated by one minus the Kaplan-Meier
estimator, where for patients without the composite event, the time to event
is censored by end of follow-up. However, whether such a composite endpoint is
an adequate safety outcome is a different question. We reiterate that any effect
on the composite may be disentangled and effects on the single components of the
composite may be analysed using the techniques discussed in Section~4.\newline
{\color{black}The \textit{hypothetical estimand} targets the effect of
  treatment on AE occurrence in the hypothetical scenario in which the
  intercurrent event treatment discontinuation would not occur. This estimand
  requires the collection of AE data only up to this event{\color{black}, under
    the assumption that both the AE hazard and the death hazard remain
    unchanged in this hypothetical situation.} The estimator used for the
  while on treatment estimand would be valid also for the hypothetical
  estimand, {\color{black}but now handling treatment discontinuation as a pure
    censoring event and not as a competing risk.}
  As th{\color{black}e} assumption is obviously {\color{black}both} untestable
  {\color{black}and rather strong}, sensitivity analyses are {\color{black}a
    minimum requirement}, when this estimand is targeted.} An even more hypothetical estimand would target the effect of treatment
on AE occurrence, assuming that neither treatment discontinuation
nor death before AE occurs. This estimand is even more hypothetical in the sense that one might be able to imagine situations where treatment continuation is
enforced, but enforcing the absence of death is much more speculative.
\newline
{\color{black}The \textit{principal stratum estimand} builds on the causal framework of
  potential primary outcomes. A potential outcome is defined for each possible
  treatment assignment, but only one potential outcome is observed, namely
  that for the actually assigned treatment. This framework is now extended to
  \emph{potential intercurrent events} and targets a causal treatment
  comparison within subpopulations defined by potential intercurrent events.
  In our example, the potential intercurrent event is treatment
  discontinuation, while alive, as a function of time and for each possible
  treatment assignment. A \emph{basic} principal stratum contains all patients
  with identical values for all potential intercurrent events. For the example
  of treatment discontinuation (but ignoring \emph{time} until discontinuation
  for ease of presentation) and two possible treatment assignments, one basic
  principal stratum is the set of all patients whose potential intercurrent
  event statuses are (yes, yes); this is the set of all patients who
  discontinue treatment under both treatment assignments. By construction,
  measuring the difference between potential outcomes on such a basic
  principal stratum yields a causal effect `adjusting' for the intercurrent
  events. Next, a principal stratum is a union of basic principal strata. A
  common example is the set of patients where the potential intercurrent
  events are unaffected by treatment --- (yes, yes) and (no, no) in the
  example above --- and the complement are all patients where the potential
  intercurrent events differ between treatments ((no, yes), (yes, no)).
  {\color{black} However, it is impossible to identify these patients in
    advance, and a post-hoc analysis{\color{black}, e.g.,} based only on those
    patients who did not stop treatment would be biased, {\color{black}one
      reason being} time-dependent confounding. For the principal stratum
    estimand causal inference methods would be required,}
  {\color{black}extending} the statistical methods presented in
  Section~4. Causal inference methods have been developed
  for the analysis of time-to-event data \citep{baker1998analysis,
    hernan18:_causal_infer}, including methods for competing events
  \citep{pka_goes_causal}{\color{black}, but practical applications may be
    subtle. For instance, a key concept in causal reasoning is that of a
    treatment regimen, determined at time~0. In our context, one may, at least
    theoretically, envisage enforcing a treatment regimen of no
    discontinuation. However, progression events (which trigger treatment
    discontinuation) are not controllable in terms of a `progression regimen',
    which is one major motivation behind the principal stratum framework
    \citep{frangakis2002principal}.}}

\section{Discussion}
The introduction of estimands provides a framework to structure the discussion about the analyses of AE in terms of specific demands/needs and appropriateness of different analyses approaches. We formulated a framework based on safety estimands within which we proposed statistical methods including methods for evidence synthesis that map the AE data to a single value. For the described estimands we have given recommendations which estimators should be used. In particular, we would like to advocate the use of time-to-event methodology for the analysis of AE data, although such a proposal is known for a quite a long time \citep{oneill1987}.\\
As discussed in Subsection 3.2, estimands of primary interest may differ between drug approval agencies and the HTA bodies in certain instances. The regulatory agencies evaluate the safety profile of a new treatment. The assessment is based on all safety data available for the new treatment which is summarized in the summary of safety. There, the safety profile of the new treatment is provided in detail showing safety data of all kinds (i.e. AEs, laboratory data, physical examination, vital signs, etc.) of all available studies with the new treatment to assess the benefit and risk of this treatment. In addition, the safety is assessed across the whole lifetime of the new treatment by assessing the required periodic safety updates. The standard approach is to use descriptive statistics such as absolute and relative frequencies. Common practice evaluates the safety of the treatment itself using a while on treatment estimand, which could also be formulated as a hypothetical estimand under relevant assumptions as stated in Section 5. The HTA bodies on the other hand are interested in the relative effect on AEs of the new treatment in comparison to a chosen comparator in the indication of interest. The estimand of interest is the treatment policy. In indications such as diabetes mellitus, studies may cover both types of estimands if treatment times and follow-up are similar in both treatment groups. In indications such as oncological diseases the follow-up of AEs is stopped at the planned end of the study treatment plus a certain number of days, which often results in considerable differences in follow-up times. In such scenarios the treatment policy estimands cannot be covered as AEs usually are not collected for subsequent therapies. A possible solution would be to plan studies that enable estimates for both estimands. All efforts should be undertaken while planning and conducting clinical trials to obtain similar follow-up times. In practice, however, this might not always be possible. Practical challenges include scenarios with patients moving on to different studies due to treatment failure. In such cases similar follow-up times cannot be guaranteed. Even if the regulatory agencies have other tasks than HTA bodies, the general objective is to establish beneficial treatments for patients. The common parts should be identified in order to harmonize both perspectives.\\
Further work, some of it under way, can be done or is required in several areas, some of which have already been mentioned. Firstly, in the present paper, we restricted ourselves to methods for analysing the time to occurrence of first AEs. Therefore, the methodologies proposed in this paper need to be revisited with a view to analysing recurrent AEs, intermittent AEs and AEs with varying severity. Secondly, our proposed methods for analysing AEs lack from adequately accounting for the occurrence of multiple, different AEs. Approaches that account for a number of types of possibly related AEs do exist \citep{Guettner2007a,Berry2004a}. Thirdly, an empirical investigation is underway to investigate in a large number of randomized controlled trials whether the different analyses of AEs lead to different decisions when comparing safety between groups. Fourthly, it is an open question what impact regulators and HTA agencies might have on the development of methods for the analysis of AE data through e.g. the ICH E9 addendum on estimands and sensitivity analysis in clinical trials. Finally, we have demonstrated that there is a gap between what would ideally be seen in benefit assessments from an HTA agency perspective and current practices of how data are collected in trials with an objective to achieve marketing authorization. It will not be easy to reconcile these two different views. However, the present article stimulated the discussion between different stakeholders. 

\section*{Acknowledgements}
The Working Group Therapeutic Research (ATF) of the German Society for Medical Informatics, Biometrics and Epidemiology (GMDS) and the Working Group Pharmaceutical Research (APF) of the German Region of the International Biometric Society (IBS-DR) have established the joint project group ``Analysis of adverse events in the presence of varying follow-up times in the context of benefit assessments''. 
We are grateful to all members of the ATF/APF project group as well as to Brenda Crowe and Ralf Bender for making valuable comments and suggestions on
the first draft of this paper. Furthermore, we would like to thank Christian Röver for his assistance in preparing the forest plot.

\bibliographystyle{unsrtnat} 
\bibliography{bibtexfile1,bibtexfile2,bibtexfile3}

\pagebreak
\begin{table}[h]
\scriptsize
\caption{Some examples from early benefit assessments with considerably different follow-up times. The ratio of the follow-up times in the fourth column is computed by dividing the follow-up time of the treatment group with shorter follow-up by the follow-up time of the group with longer follow-up. The dossier assessments can be obtained from https://www.iqwig.de.}
$ $ \newline
\begin{tabular}{l|l|l|l}
 \hline
Dossier evaluation	& Intervention &	Control	& Ratio of follow-up times \\
\hline
Oncology &   \multicolumn{2}{c|}{Median follow-up + safety follow-up} & \\	 	 	
A14-48  prostate	& 16.6 months + 28 days	& 4.6 months + 28 days &	31\% \\
A15-17  lung	& 336 + 28 days &	105 + 28 days &	37\% \\
A15-33  melanoma	& 168 + 90 days &	63 + 90 days &	59\% \\
A16-04  mantle cell lymphoma &	14.4 months + 30 days &	3.0 months + 30 days &	26\% \\
\hline
Hepatitis C	 &  \multicolumn{2}{c|}{Planned follow-up + safety follow-up} & \\	 	
A14-44 &	8 - 12 weeks + 30 days	& 24, 28 or 48 weeks + 30 days	& 23\% - 57\% \\
A16-48 &	12 weeks + 30 days	& 24 weeks + 30 days &	57\% \\
\hline
\end{tabular}
\end{table}

\begin{center}
\begin{figure}[h]
\centering
\includegraphics[scale=0.8]{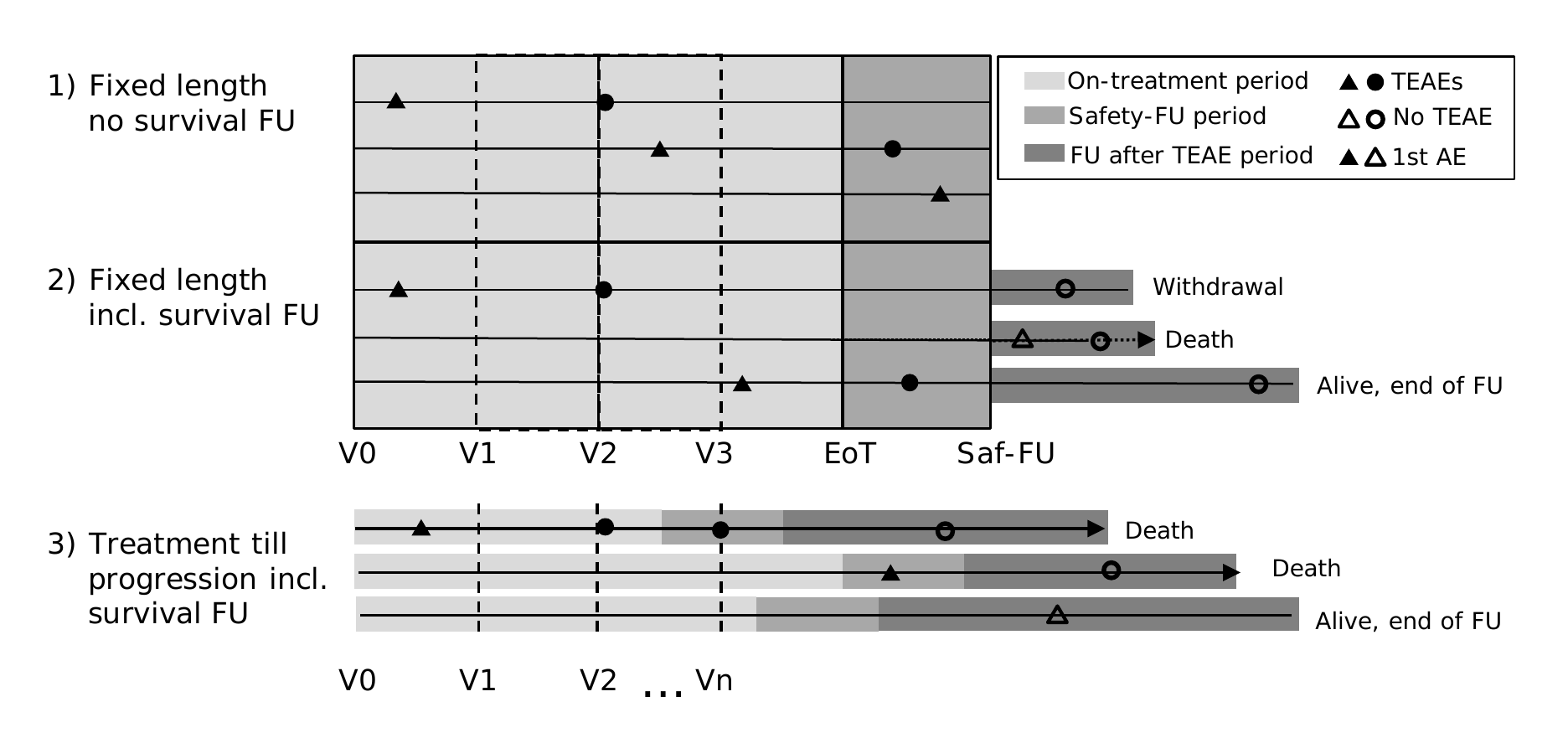}
\caption{Description of different scenarios for typical AE follow-up (FU) in clinical trials (TEAEs: treatment emergent AEs (marked by bold symbols); EoT: end of treatment; Saf-FU: safety follow-up; V0: visit at the beginning of the trial; V1,...,Vn: visits during treatment). First occurrences of AEs are marked by triangles.}
\end{figure}
\end{center}

\pagebreak

\begin{center}
\begin{figure}[h]
\centering
\includegraphics[scale=0.5]{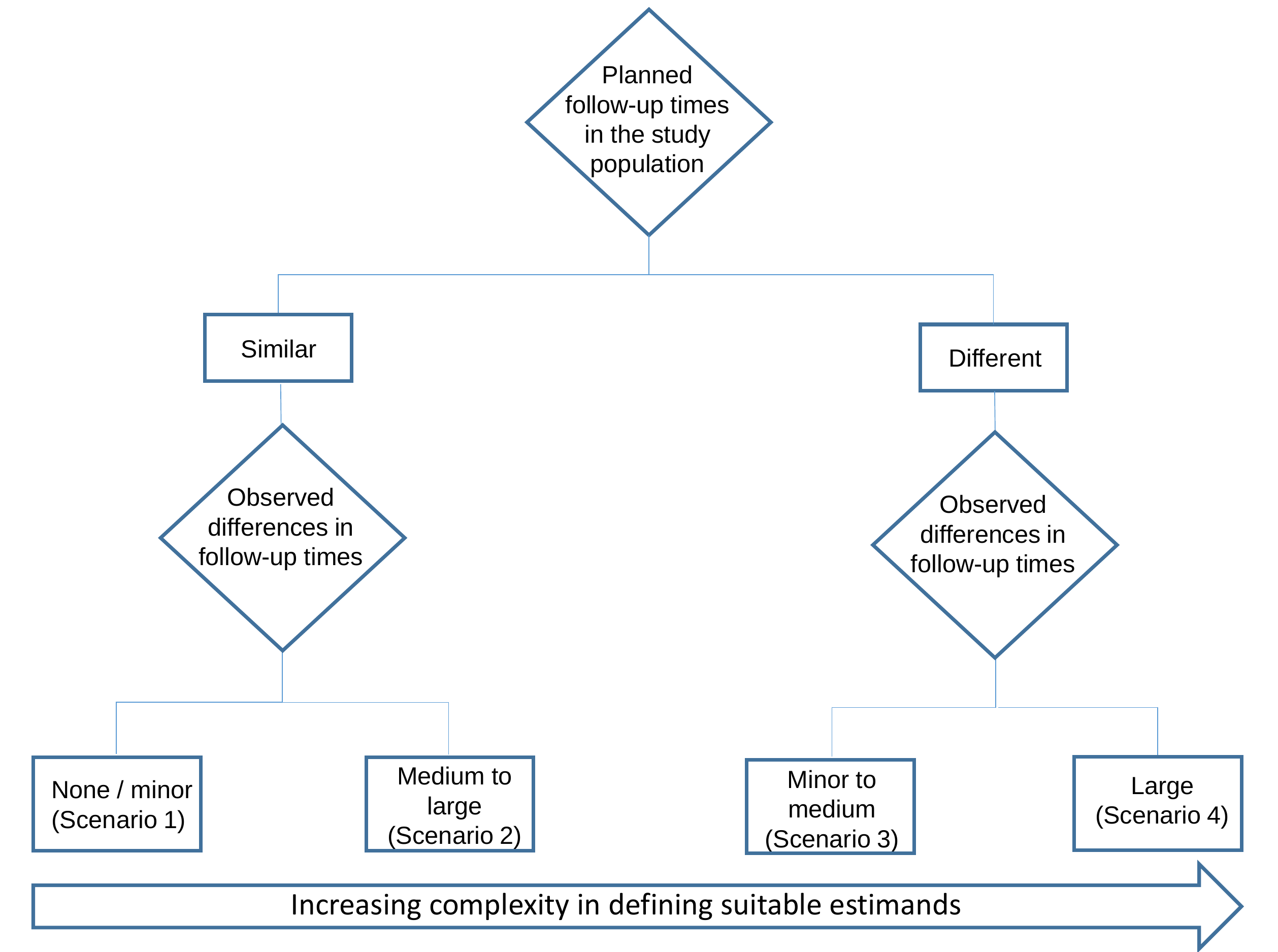}
\caption{Flow chart displaying four different scenarios across indications for the consideration of safety estimands in an HTA system.}
\end{figure}
\end{center}

\begin{center}
\begin{figure}[h]
\centering
\includegraphics[scale=0.7]{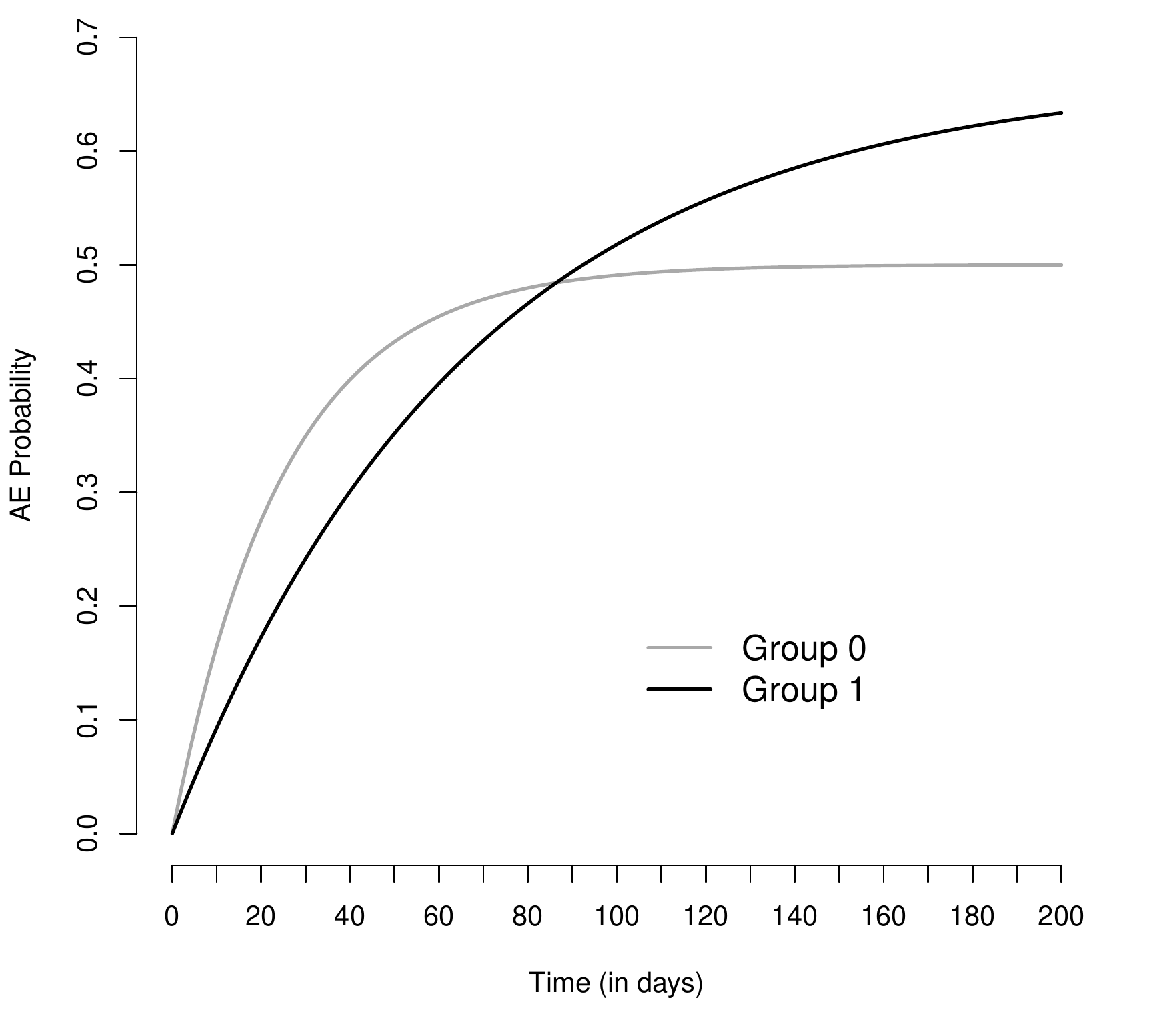}
\caption{Cumulative AE probabilities for two groups and constant hazards. Although in group 1 the AE hazard is lower compared to group 0, the cumulative AE probability in group 1 is eventually greater than in group 0.}
\end{figure}
\end{center}

\begin{center}
\begin{figure}[h]
\centering
\includegraphics[scale=0.75]{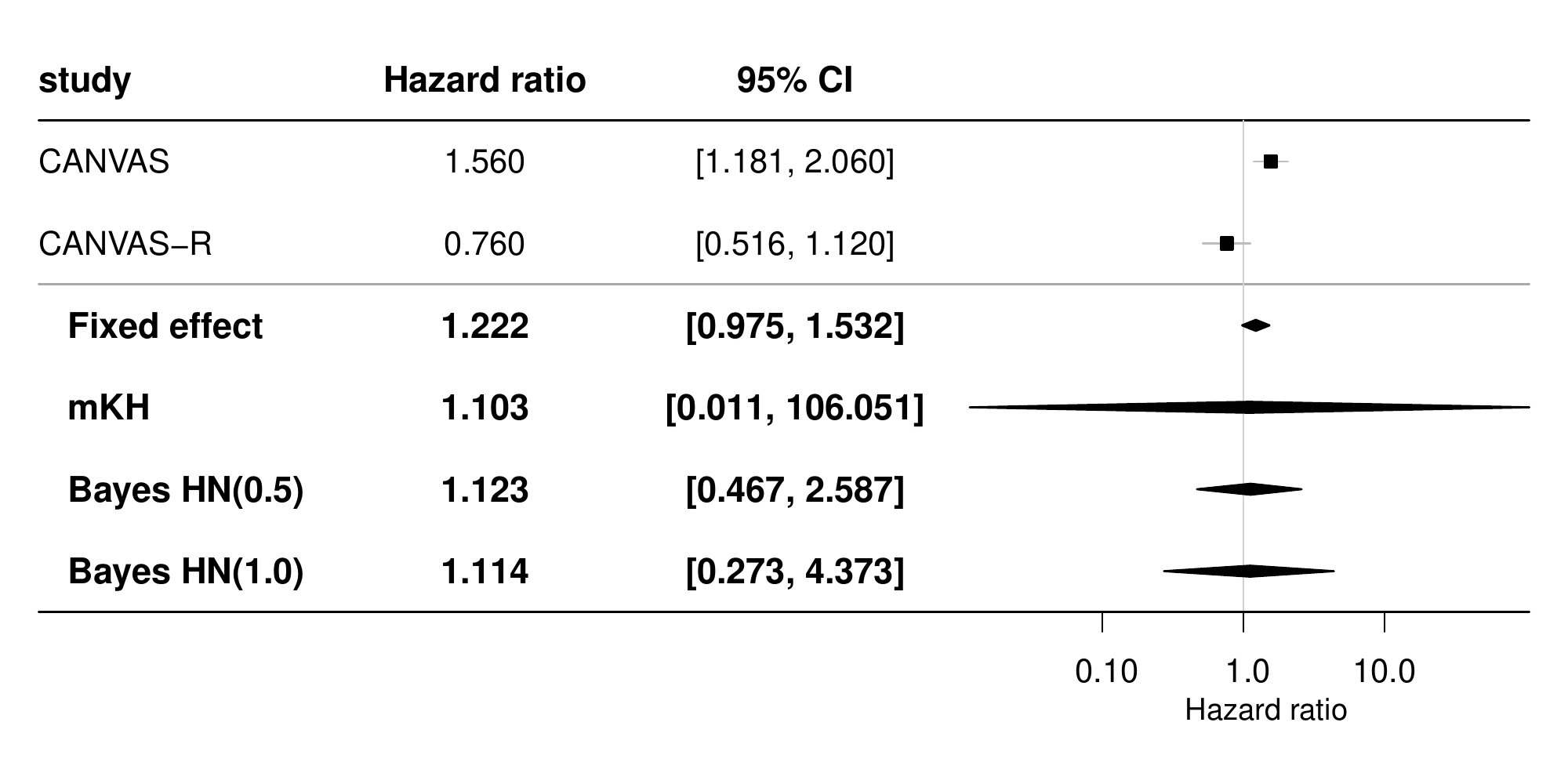}
\caption{Illustrating example for meta-analyses. Forest plot of hazard ratios for low trauma fractures as observed in CANVAS and CANVAS-R with 95\% CIs and four combined hazard ratios from a fixed-effect meta-analysis, modified Knapp-Hartung (mKH) meta-analysis and Bayesian random-effects meta-analysis with two half-normal (HN) priors for the heterogeneity parameter $\tau$.}
\end{figure}
\end{center}

\end{document}